\documentclass[pra,aps,twocolumn,superscriptaddress,showpacs]{revtex4}
\usepackage{mathrsfs}
\usepackage{amssymb}
\usepackage{bbm}
\usepackage{amsmath}
\usepackage{graphicx}
\usepackage{epsfig,amsmath}

\begin{document}

\title
{Quantum dynamical algebra in exactly solvable one-dimensional
 potentials}
 \author{Ming-Guang Hu}
\email{freedom@mail.nankai.edu.cn} \affiliation{Theoretical Physics
Division, Chern Institute of Mathematics, Nankai University, Tianjin
300071, P. R. China}
\author{Jing-Ling Chen}
 \email{chenjl@mail.nankai.edu.cn}
\affiliation{Theoretical Physics Division, Chern Institute of
Mathematics, Nankai University, Tianjin 300071, P. R. China}

\date{\today}
\begin{abstract}
We mainly explore the linear algebraic structure like $SU(2)$ or
$SU(1,1)$ of the shift operators for some one-dimensional exactly
solvable potentials in this paper. During such process, a set of
method based on original diagonalizing technique is presented to
construct those suitable operator elements, $J_0$, $J_\pm$ that
satisfy $SU(2)$ or $SU(1,1)$ algebra. At last, the similarity
between radial problem and one-dimensional potentials encourages us
to deal with the radial problem in the same way. And the
corresponding algebra turns to approach $SU(1,1)$ algebra but for
$J_0\neq J_0^\dagger$, $J_+^\dagger\neq J_-$.
\end{abstract}
  \maketitle


\section{Introduction}
Exactly solvable potential especially including one-dimensional or
spherically symmetric ones are playing the indispensable role in
condensed matter, biophysics, nuclear physics, quantum optics, and
solid-state physics, etc. Those familiar potentials embrace, for
instance, the typical harmonic oscillator potential, conventional
Coulomb potential, one dimensional Morse potential
\cite{Morse(1929)}, the Rosen-Morse potential \cite{Rosen(1932)},
P\"{o}schl-Teller potential \cite{Poschl(1933)}, the Hulth\'{e}n
potential, the Kratzer's molecular potential, and the famous Yukawa
potential, etc. Thereinto, the Morse potential and Kratzer's
molecular potential are utilized to describe the anharmonicity and
bond dissociation of diatomic molecules. Another noticeable
potential with short-range properties is P\"{o}schl-Teller
potential, of which the generalized coherent states
\cite{Crawford(1998)}, nonlinear properties \cite{Chen(1998),
Quesne(1999)}, and supersymmetric extension \cite{Diaz(1999)} have
been extensively studied. Additionally, in supersymmetric quantum
mechanics (SUSYQM), the shaped invariant potentials have also been
mooted
\cite{Gendenstein(1983),Dutt(1986),Cooper(1988),Khare(1988),Dabrowska(1988)}.
In particular, a large class of such potential is the Natanzon class
\cite{Natanzon(1979),Fukui(1993)}.

The preference to deal with those potentials in modern quantum
mechanics adopt the abstract formulation and they stress the special
nature of wave mechanics. However, the machinery of wave mechanics
such as choice of coordinate system, separation of variables,
boundary conditions, single-valuedness can obscure the underlying
quantum mechanical principles and complicate the analysis. In this
way, the operator methods which mainly consist of noncommutative
algebra and the shift operator factorization to some extent can
avoid these flaws. 
Algebraic methods \cite{Kamran(1990),Celeghini(1985)} exploring the
underlying Lie symmetry and its associated algebra have been widely
used to study many of these exactly solvable potentials, for
instance, Darboux transformation, Infeld-Hull transformation
\cite{Infeld(1951)}, Mielnick facorization \cite{Mielnik(1984)},
SUSY quantum mechanics, inverse scattering theory
\cite{Hoppe(1992)}, and intertwining technique \cite{Diaz(1999)}. On
the other hand, by exploiting an underlying q-deformation quantum
algebraic symmetry, Spiridonov \cite{Spiridonov(1992)} has shown how
a finite differential equation can generate in the limiting cases a
set of exactly solvable potentials.

Recently, Chen et al. \cite{Chen(1998)} applied nonlinear
deformation algebra (NLDA) introduced by Delbecq and Quesne
\cite{Delbecq(1993)} to a physical system with P\"{o}schl-Teller
potential, and further they obtained the $SU(1,1)$ algebra from
$NLDA$ naturally \cite{Chen(1998),Quesne(1999)}. Then by using the
similar method in \cite{Ge(2000)} a unified approach that emphasized
on constructing the shift operators of exactly one-dimensional
solvable potentials was given without pointing out concomitant
algebraic structures. And it is the main purpose of this paper to
find a systematical method to construct the $SU(2)$ or $SU(1,1)$
algebra for these potentials. Additionally, for the similarity
between the radial part of the spherically symmetric potentials  and
one-dimensional potential, we also have a try to explore the method.

This paper is arranged as follows: In Sec. \ref{sec-Quan}, we first
briefly reviewed the PT problem discussed in
\cite{Chen(1998),Quesne(1999)} and then gave out a systematic method
to construct the linear algebra for one-dimensional exactly solvable
potentials; In Sec. \ref{sec-mor}, some one-dimensional exactly
solvable potentials were discussed with the use of the method; At
last, we  in the same measure discussed the radial problem in Sec.
\ref{sec-rad}.

\section{Quantum dynamic algebra of one-dimensional
potential}\label{sec-Quan}
\subsection{Brief Review} The nonlinear deformation algebra (NLDA) as
noted in \cite{Delbecq(1993)} generated by three operators
$b_0=b_0^\dagger$, and $b_-=(b_+)^\dagger$ satisfying
\begin{eqnarray}
[b_0, b_-]&=&-b_-g(b_0), \quad [b_0, b_+]= g(b_0)b_+, \nonumber\\
&&[b_-, b_+]=f(b_0).\label{eq-NLDA}
\end{eqnarray}
which can be realized for the P\"{o}schl-Teller (PT) potential
pointed out by Chen et al. in \cite{Chen(1998)}. Further, such NLDA
of PT potential can be transformed into a $SU(1,1)$ algebra
demonstrated in \cite{Chen(1998),Quesne(1999)}. The results are
briefly reviewed as follows.

The Hamiltonian is $H=p^2/2m+V(x)$ with $V(x)=V_0/\cos^2(kx)$, and
$V_0=\varepsilon\nu(\nu-1)$ accompanying by
$\varepsilon=\hbar^2k^2/2m$. After defining the generalized
coordinate $X$ and momentum $P$,\[X=\sin(kx),\quad
P=\frac{1}{2}k\{\cos(kx),p\},\] it found out the shift operator in
Eq. (\ref{eq-NLDA})
\begin{eqnarray}
b_0&=&H,\nonumber\\
b_-&=&\frac{1}{2\varepsilon}\left[X\left(\varepsilon+2\sqrt{\varepsilon
H}\right)+\frac{i\hbar}{m}P\right],\nonumber\\
b_+&=&-\frac{1}{2\varepsilon}\left[X\left(\varepsilon-2\sqrt{\varepsilon
H}\right)+\frac{i\hbar}{m}P\right]\frac{\varepsilon+\sqrt{\varepsilon
H}}{\sqrt{\varepsilon H}}.
\end{eqnarray}
and
\begin{eqnarray*}
g(H)&=&-\varepsilon+2\sqrt{\varepsilon H},\\
f(H)&=&1+2\sqrt{H/\varepsilon}+\frac{\nu(\nu-1)}{\sqrt{H/\varepsilon}(\sqrt{H/\varepsilon}-1)}
\end{eqnarray*}
Thus it realized the NLDA. Further, through redefining the three
generator elements, that is
\begin{equation*}
J_0=\sqrt{H/\varepsilon},~~J_-=\left(\frac{\sqrt{H/\varepsilon}}{\sqrt{H/\varepsilon}+1}\right)^{\frac{1}{2}}b,~~J_+=J_-^\dagger
\end{equation*}
they satisfied the $SU(1,1)$ algebra
\[[J_0,J_\pm]=\pm J_\pm,\quad [J_+,J_-]=-2J_0.\]

Therefore, it turned the nonlinear algebra into a linear dynamical
algebra for such special PT potential that possesses a closed
operator sets $\{H, X, P\}$. Such procedure also gives a suggestion
that for an arbitrary one-dimensional potential, if finding its
closed operator sets $\{H, X, P\}$, we can immediately construct the
shift operators and then tune them into a linear algebra. The method
to find the closed operator sets has been discussed in
\cite{Ge(2000)}, and in the next subsection we emphasize the method
to tune them into a linear algebra.

\subsection{The main method} For a given quantum system with
Hamiltonian $H$, if there are generalized coordinate operator $Q$
and generalized momentum operator $P$ satisfying the following
commutation relations as described in \cite{Ge(2000)}
\begin{eqnarray*}
[H,Q]&=&Q\Theta_1(H)+P\Pi_1(H),\\
~[H,P]&=&Q\Theta_2(H)+P\Pi_2(H),
\end{eqnarray*}
where $\Theta_i(H)$ and $\Pi_i(H)$ are functions about operator $H$,
it is then able to find out the shift operators for the energy
eigenstates by using the matix-diagonalizing technique. That is,
given
\begin{equation}\label{eq-HQP}
[H,(Q,P)]=(Q,P)\left(\begin{matrix} \Theta_1&\Theta_2\\\Pi_1&\Pi_2
\end{matrix}\right)
\end{equation}
diagonalizing the matrix in the right obtain a transformation matrix
$S$ and leaves a diagnolized matrix $\Lambda$. Thus the Eq.
(\ref{eq-HQP}) can be written as
\begin{equation*}
[H, (Q,P)]=(Q,P)S\Lambda S^{-1}
\end{equation*}
with \[\Lambda=\left(\begin{matrix} -\Omega_1(H)&0\\0&\Omega_2(H)
\end{matrix}\right).\]
We can define the shift operators $S_1$ and $S_2$ to be
\[(S_1,S_2)=(Q,P)S\]
which satisfy
\begin{equation}\label{eq-HSS}
H(S_1,S_2)=(S_1,S_2)\left(\begin{matrix}
H-\Omega_1(H)&0\\0&H+\Omega_2(H)
\end{matrix}\right)
\end{equation}
Further, for a real function $F(H)$, holomorphic in the neighborhood
of zero, it is easy to derive the following identity
\begin{equation}\label{eq-Fss}
F(H)(S_1,S_2)=(S_1,S_2)\left(\begin{matrix}
F(H-\Omega_1)&0\\0&F(H+\Omega_2)
\end{matrix}\right)
\end{equation}

 From the Eqs. (\ref{eq-HSS}) and (\ref{eq-Fss}) we can always find
such operator function $b_0(H)$ that satisfy
\begin{equation}\label{eq-b0}
b_0(H)(S_1,S_2)=(S_1,S_2)\left(\begin{matrix} b_0(H)-1&0\\0&b_0(H)+1
\end{matrix}\right)
\end{equation}
which have a more familiar form as
\begin{equation}\label{eq-bS}
[b_0,S_1]=-S_1,\quad [b_0,S_2]=S_2.
\end{equation}
Such commutation relation implies a freedom that
\begin{equation}\label{eq-free2}
b_0(H)\longrightarrow b_0(H)+\xi_0
\end{equation}
with $\xi_0$ a constant keeps the Eq. (\ref{eq-bS}) unchangeable.
This freedom would be explored to select out the Lie algebraic
element $J_0$. Additionally, the identity (\ref{eq-Fss}) evolved
into
\begin{equation}\label{eq-Gss}
G(b_0)(S_1,S_2)=(S_1,S_2)\left(\begin{matrix} G(b_0-1)&0\\0&G(b_0+1)
\end{matrix}\right)
\end{equation}
One  noticeable point is that there is a freedom of multiplying a
function $\xi_i(b_0)$ on $S_1$ and $S_2$, that is
\begin{eqnarray}
S_1&\longrightarrow& \xi_1(b_0)S_1\  \text{or}\
S_1\xi_1(b_0),\nonumber\\S_2&\longrightarrow& \xi_2(b_0)S_2\
\text{or}\ S_2\xi_2(b_0)\label{eq-free1}
\end{eqnarray}
which still satisfy the  Eqs. (\ref{eq-HSS}) and (\ref{eq-Fss}).
This freedom can be used to guarantee finding the Lie algebraic
elements $J_+$ and $J_-$ from $S_1$ and $S_2$.

As for the constructed shift operators $S_1$ and $S_2$, generally
they are not Hermite conjugate with each other but satisfy
\begin{equation}\label{eq-sseta}
S_1^\dagger=S_2 \eta(b_0).
\end{equation}
which makes it easy to substitute
\begin{equation}\label{eq-bs}
b=S_1,\quad b^\dagger=S_1^\dagger=S_2\eta(b_0)
\end{equation}
 with
$b^\dagger=(b)^\dagger$. For the purpose of constructing an algebra
structure, we can define
\begin{equation}\label{eq-JJde}
J_0=b_0+\xi_0,\quad J_+=b^\dagger\xi(b_0),\quad J_-=\xi(b_0)b
\end{equation}
which have naturally satisfied \[[J_0,J_+]=J_+,\quad
[J_0,J_-]=-J_-,\quad J_+=J_-^\dagger,\] and $\xi_0$, $\xi(b_0)$ need
 finally determined by the case
\begin{equation}
[J_+,J_-]=\pm2 J_0,
\end{equation}
where `+' corresponds to $SU(2)$ while `-' to $SU(1,1)$.

 The commutation relation from the Eqs. (\ref{eq-Gss}) and (\ref{eq-sseta})
 is reduced into
\begin{eqnarray*}
[J_+,
J_-]&=&b^\dagger\xi(b_0)^2b-\xi(b_0)bb^\dagger\xi(b_0)\nonumber\\
&=&S_2S_1\eta(b_0-1)\xi(b_0-1)^2-S_1S_2\eta(b_0)\xi(b_0)^2
\end{eqnarray*}
which would be treated in three cases.
\begin{itemize}
\item The $\eta(b_0)=Constant$, but $[S_1, S_2]=f(b_0)\neq Constant$
case. We can simply put, $\xi(b_0)=\xi(b_0-1)=Constant$, and thus
\begin{equation}\label{eq-case1}
[J_+,J_-]=[S_2,S_1]\eta\xi=f(b_0)\eta\xi=\pm2J_0
\end{equation}
\item The $\eta(b_0)=Constant$, and $[S_1, S_2]=Constant$
case. We must choose $\xi(b_0)=f(b_0)\neq Constant$ to satisfy
\begin{eqnarray}
[J_+,J_-]&=&[S_2S_1f(b_0-1)-S_1S_2f(b_0)]\xi^2\nonumber\\
&=&\pm2J_0\label{eq-case2}
\end{eqnarray}
\item The $\eta(b_0)\neq Constant$ case.
We can choose $\xi(b_0)\neq Constant$ to satisfy
\begin{equation}\label{eq-etaxi}
\eta(b_0-1)\xi(b_0-1)^2=\eta(b_0)\xi(b_0)^2=g(b_0),
\end{equation}
and thus
\begin{equation}\label{eq-case3}
[J_+,J_-]=[S_2,S_1]g(b_0)=\pm2J_0.
\end{equation}
\end{itemize}
In this way, the Eqs. (\ref{eq-case1}), (\ref{eq-case2}),
(\ref{eq-etaxi}), and (\ref{eq-case3}) can be finally used to derive
out the algebra elements $J_0$, $J_\pm$.

One simple example concerns the one-dimensional harmonic oscillator
with the Hamiltonian, $H=\frac{1}{2}(x^2+p^2)$, where $x$ is the
coordinate operator and $p=-i(d/dx)$ is the momentum operator with
commutator given by $[x,p]=i$. For the commutation relations
$[H,x]=-ip$ and $[H,p]=ix$, they as noted in \cite{Ge(2000)} can be
succinctly written as
\begin{equation*}
[H, (x,p)]=(x,p)\left(\begin{matrix} 0&i\\-i&0
\end{matrix}\right).
\end{equation*}
Finding out its shift operators $a$ and $a^\dagger$, it gives
\begin{equation*}
[H, (a,a^\dagger)]=(a,a^\dagger)\left(\begin{matrix} -1&0\\
0&1
\end{matrix}\right)
\end{equation*}
in which $a\equiv (1/\sqrt{2})(x+ip)$ and
$a^\dagger\equiv(1/\sqrt{2})(x-ip)$ with
\begin{equation}\label{eq-harm1}
aa^\dagger=H+\frac{1}{2},\qquad [a,a^\dagger]=1.
\end{equation}
 Thus, from Eqs. (\ref{eq-b0}) and
(\ref{eq-bs}) we know
\begin{equation}\label{eq-harm2}
J_0=H,\quad b=a,\quad b^\dagger=a^\dagger.
\end{equation}
It belongs to the second case. Inserting Eqs. (\ref{eq-harm1}) and
(\ref{eq-harm2}) into Eq. (\ref{eq-case2}), it deduces
\begin{equation*}
(H-\frac{1}{2})\xi(H-1)^2-(H+\frac{1}{2})\xi(H)^2=-2H.
\end{equation*}
Obviously, by choosing $\xi(H)^2=H+1/2$, the above equation would
hold. Therefore, the one-dimensional harmonic oscillator has a
$SU(1,1)$ algebra with
\begin{equation*}
J_0=H,\  J_+=a^\dagger \sqrt{H+1/2},\  J_-=\sqrt{H+1/2}\ a.
\end{equation*}

\section{Quantum dynamical algebra for more
potentials}\label{sec-mor}
 As pointed out in \cite{Ge(2000)}, the
shift operators of a class of solvable one-dimension potentials can
be found by using the matrix-diagonalizing technique. In the
following, we will continue to explore such technique to display
that all these potentials possess either a $SU(1,1)$ or a $SU(2)$
algebra. To this end, we begin with the following Hamiltonian:
\begin{equation*}
H=X(x)\frac{d^2}{dx^2}+V(x),
\end{equation*}
where $X(x)$ is an arbitrary function of position and $V(x)$ is an
arbitrary potential. Define
\begin{eqnarray*}
P(x,p)&=&Y(x)\frac{d}{dx}+Z(x);
\end{eqnarray*}
with arbitrary functions $Y(x)$ and $Z(x)$ to be determined. The
following commutation relations hold
\begin{eqnarray}
\big[H,P(x)\big]&=&Q(x)(\beta H+1)+\alpha P+\gamma H,\nonumber\\
\big[H,Q(x)\big]&=&2\lambda P+\nu Q(x)+\tau.\label{eq-HPQ}
\end{eqnarray}
with
\begin{eqnarray}
 Q(x)&=&\frac{1}{1+\beta
V(x)}\bigg[X(x)Z''(x)-\gamma V(x)-\alpha
Z(x)\nonumber\\&&-Y(x)V'(x)\bigg].
\end{eqnarray}
Concomitantly, $X(x)$, $V(x)$, $Y(x)$ and $Z(x)$ satisfy
\begin{eqnarray}
&&X(x)[Y''(x)+2Z'(x)]=\alpha Y(x),\nonumber\\
&&2X(x)Y'(x)-X'(x)Y(x)=[\beta Q(x)+\gamma]X(x),\label{eq-cosis}\\
&&X(x)Q'(x)=\lambda Y(x),\nonumber\\
&&-2\lambda Z(x)+X(x)Q''(x)=\nu Q(x)+\tau.\nonumber
\end{eqnarray}
In the construction, we should bear in mind that $Y(x)$, $Z(x)$,
$Q(x)$, and $\alpha$, $\beta$, $\gamma$, $\lambda$, $\nu$, $\tau$
are all determined by $X(x)$ and $V(x)$ when we construct the shift
operators for a given Hamiltonian. Conversely, if we proceed with a
systematic search for exactly solvable potentials based on this
method, we can have as many degrees of freedom as there are free
parameters and functions, namely, $X(x)$, $Y(x)$, $Z(x)$, $Q(x)$,
and $\alpha$, $\beta$, $\gamma$, $\lambda$, $\nu$, $\tau$.

\subsection{The $X(x)=-1$, $Y(x)=i$ case} By substituting
into Eq. (\ref{eq-cosis}), we solved
\begin{equation}
Q(x)=-i\lambda x+c_1,\quad Z(x)=-\frac{i}{2}\alpha x+c_2,
\end{equation}
and
\begin{equation}
V(x)=\frac{1}{2}\left(\lambda+\frac{\alpha^2}{2}\right)x^2-(c_1+\alpha
c_2)ix+c_3,
\end{equation}
with $\beta=0$, $\gamma=0$, $\nu=-\alpha$ and $\tau=-c_1\nu-2\lambda
c_2$. This is the harmonic oscillator when the coefficient
$\lambda+\alpha^2/2>0$. For the requirement of Hermitian operators
$H$ and $P$, it requires the parameters $\alpha$, $c_1$ being
imaginary numbers while $\lambda$, $c_2$, $c_3$ real numbers.

The closed operator set $\{H, \tilde{Q}, \tilde{P}\}$ satisfy the
commutation relations
\begin{equation}
[H, \tilde{Q}]=-\alpha \tilde{Q}+2\lambda \tilde{P}, ~~~~[H,
\tilde{P}]=\tilde{Q}+\alpha \tilde{P}.
\end{equation}
 or more succinctly as
\begin{equation}
[H, (\tilde{Q},\tilde{P})]=(\tilde{Q},\tilde{P})\left(\begin{matrix}
-\alpha&1\\2\lambda&\alpha
\end{matrix}\right)
\end{equation}
in which
\begin{eqnarray}
\tilde{Q}&=&Q+\alpha\frac{2\lambda c_2+\nu
c_1}{\alpha^2+2\lambda}=-i\lambda x+c_1+\alpha\frac{2\lambda c_2+\nu
c_1}{\alpha^2+2\lambda},\nonumber\\
\tilde{P}&=&P-\frac{2\lambda c_2+\nu
c_1}{\alpha^2+2\lambda}=i\frac{d}{dx}-\frac{i}{2}\alpha
x+\frac{\alpha^2c_2-\nu c_1}{\alpha^2+2\lambda}.
\end{eqnarray}
It can be easily diagonalized to give the shift operators $S_1$ and
$S_2$ which satisfied the commutation relation
\begin{equation}
[H, (S_1,S_2)]=(S_1,S_2)\left(\begin{matrix}
-\sqrt{\alpha^2+2\lambda}&0\\0&\sqrt{\alpha^2+2\lambda}
\end{matrix}\right)
\end{equation}
with
\begin{eqnarray}
S_1&=&\frac{1}{\alpha-\sqrt{\alpha^2+2\lambda}}\tilde{Q}+\tilde{P},\nonumber\\S_2&=&\frac{1}{\alpha+\sqrt{\alpha^2+2\lambda}}\tilde{Q}+\tilde{P}
\end{eqnarray}
which satisfy the relation $(S_1)^\dagger=S_2$ and
\begin{equation}
[S_1,S_2]=\sqrt{\alpha^2+2\lambda}.
\end{equation}
So it belongs to the second case. Thus, we can obtain a $SU(1,1)$
algebra according to the second treatment after defining
\begin{equation}
 J_0=H/\sqrt{\alpha^2+2\lambda},
\end{equation}
\begin{equation}
J_-=\frac{\sqrt{J_0+1/2}}{(\alpha^2+2\lambda)^{1/4}}b,\quad
J_+=b^\dagger\frac{\sqrt{J_0+1/2}}{(\alpha^2+2\lambda)^{1/4}}
\end{equation}

\subsection{The $X(x)=-1$, $Y(x)=x$ case}
By substituting into Eq. (\ref{eq-cosis}), we solved
\begin{equation}
Q(x)=-\frac{\lambda}{2} x^2+c_1,\quad Z(x)=-\frac{\alpha}{4}x^2+c_2,
\end{equation}
and
\begin{equation}\label{eq-radia}
V(x)=\frac{1}{16}\left(\alpha^2+2\lambda\right)x^2+\frac{c_3}{x^2}+\frac{1}{2}\left(\frac{\alpha}{2}-\alpha
c_2-c_1\right),
\end{equation}
with $\beta=0$, $\gamma=2$, $\nu=-\alpha$ and
$\tau=(1-2c_2)\lambda+\alpha c_1$. this gives us the radial harmonic
oscillator potential.

The closed operator set $\{H, \tilde{Q}, \tilde{P}\}$ satisfy the
commutation relations
\begin{equation}
[H, (\tilde{Q},\tilde{P})]=(\tilde{Q},\tilde{P})\left(\begin{matrix}
-\alpha&1\\2\lambda&\alpha
\end{matrix}\right)
\end{equation}
in which
\begin{eqnarray*}
\tilde{Q}&=&Q+\frac{4\lambda}{\alpha^2+2\lambda}H-\frac{\alpha}{\alpha^2+2\lambda}(\lambda+\alpha c_1-2\lambda c_2)\nonumber\\
&=&-\frac{\lambda}{2}
x^2+c_1+\frac{4\lambda}{\alpha^2+2\lambda}H-\frac{\alpha}{\alpha^2+2\lambda}(\lambda+\alpha
c_1-2\lambda c_2)\\
\tilde{P}&=&x\frac{d}{dx}-\frac{\alpha}{4}x^2+c_2+\frac{2\alpha}{\alpha^2+2\lambda}H+\frac{1}{\alpha^2+2\lambda}\nonumber\\
&&\times(\lambda+\alpha c_1-2\lambda c_2).
\end{eqnarray*}
The shift operators $S_1$ and $S_2$ which satisfied the commutation
relation
\begin{equation}
[H, (S_1,S_2)]= (S_1,S_2)\left(\begin{matrix}
-\sqrt{\alpha^2+2\lambda}&0\\0&\sqrt{\alpha^2+2\lambda}
\end{matrix}\right),
\end{equation}
and
\begin{equation*}
[S_1,S_2]=16\frac{\lambda}{\sqrt{\alpha^2+2\lambda}}\left[H-\frac{1}{2}\left(\frac{\alpha}{2}-\alpha
c_2-c_1\right)\right],
\end{equation*}
 with
\begin{eqnarray*}
S_1&=&\tilde{Q}+\left(\alpha-\sqrt{\alpha^2+2\lambda}\right)\tilde{P},\nonumber\\S_2&=&\tilde{Q}+\left(\alpha+\sqrt{\alpha^2+2\lambda}\right)\tilde{P}
\end{eqnarray*}

The shift operator $S_1$ and $S_2$ have
\begin{eqnarray*}
S_1^\dagger&=&\left(1+\frac{\alpha^2-\alpha\sqrt{\alpha^2+2\lambda}}{\lambda}\right)\tilde{Q}+(-\alpha+\sqrt{\alpha^2+2\lambda})\tilde{P}\nonumber\\
&=&S_2\left(1+\frac{\alpha^2-\alpha\sqrt{\alpha^2+2\lambda}}{\lambda}\right)
\end{eqnarray*}
Therefore we can define
\begin{eqnarray*}
&b_0=\frac{H}{\sqrt{\alpha^2+2\lambda}},\quad b=S_1,&\\
&b^\dagger=S_2\left(1+\frac{\alpha^2-\alpha\sqrt{\alpha^2+2\lambda}}{\lambda}\right).&
\end{eqnarray*}
Thus it belongs to the first case. According to the first treatment,
we can define again
\begin{eqnarray}
J_0&=&b_0-\frac{1}{2\sqrt{\alpha^2+2\lambda}}\left(\frac{\alpha}{2}-\alpha
c_2-c_1\right)\\
J_+&=&\frac{1}{2\sqrt{2}}\left(\alpha^2+\lambda-\alpha\sqrt{\alpha^2+2\lambda}\right)^{-1/2}b^\dagger,\\
J_-&=&\frac{1}{2\sqrt{2}}\left(\alpha^2+\lambda-\alpha\sqrt{\alpha^2+2\lambda}\right)^{-1/2}b,
\end{eqnarray}
which satisfy the $SU(1,1)$ algebra \[[J_+,J_-]=-2J_0.\]

\subsection{The $X(x)=-1$, $Y(x)=ae^{cx}+be^{-cx}$ case}

By substituting into Eq. (\ref{eq-cosis}), we solved
\begin{eqnarray}
Q(x)&=&-\frac{\lambda}{c}\left(ae^{cx}-be^{-cx}\right)+c_1,\\
Z(x)&=&-\frac{\alpha+c^2}{2c}\left(ae^{cx}-be^{-cx}\right)+c_2,
\end{eqnarray}
and
\begin{equation}\label{eq-radia}
V(x)=c_3\left(ae^{cx}+be^{-cx}\right)^{-2}+\frac{(\alpha+c^2)^2+c^4+2\lambda}{4c^2},
\end{equation}
with $\beta=-2c^2/\lambda$, $\gamma=2c^2c_1/\lambda$,
$\nu=-\alpha-2c^2$, $\tau=-2\lambda c_2-\nu c_1$,  and $c_1+\alpha
c_2=0$. This gives us the second P\"{o}schl-Teller potential.

The closed operator set $\{H, \tilde{Q}, \tilde{P}\}$ satisfy the
commutation relations\begin{equation} [H,
(\tilde{Q},\tilde{P})]=(\tilde{Q},\tilde{P})\left(\begin{matrix}
-\alpha-2c^2&1-\frac{2c^2}{\lambda}H\\2\lambda&\alpha
\end{matrix}\right)
\end{equation}
in which
\begin{eqnarray}
\tilde{Q}=Q-c_1&=&-\frac{\lambda}{c}\left(ae^{cx}-be^{-cx}\right)\\
\tilde{P}=P+\frac{c_1}{\alpha}&=&\left(ae^{cx}+be^{-cx}\right)\frac{d}{dx}-\frac{\alpha+c^2}{2c}\nonumber\\
&&\times\left(ae^{cx}-be^{-cx}\right)
\end{eqnarray}
The shift operators $S_1$ and $S_2$ satisfy the commutation relation
\begin{eqnarray}
[H,S_1]&=&S_1\left[-c^2-\sqrt{(\alpha+c^2)^2+2\lambda-4c^2
H}\right],\\
~[H,S_2]&=&S_2\left[-c^2+\sqrt{(\alpha+c^2)^2+2\lambda-4c^2
H}\right].
\end{eqnarray}
with
\begin{eqnarray}
S_1&=&-\tilde{Q}\frac{1}{2\lambda}\left[\alpha+c^2+\sqrt{(\alpha+c^2)^2+2\lambda-4c^2
H}\right]+\tilde{P},\nonumber\\
S_2&=&-\tilde{Q}\frac{1}{2\lambda}\left[\alpha+c^2-\sqrt{(\alpha+c^2)^2+2\lambda-4c^2
H}\right]+\tilde{P}.\nonumber\\
\end{eqnarray}
and
 \begin{eqnarray*}
& [S_1,S_2]=-8abc^2b_0,&\\
&S_2^\dagger=S_1\left[\frac{2c^2}{\sqrt{(\alpha+c^2)^2+2\lambda-4c^2
H}}-1\right].&
\end{eqnarray*}

From the Eq. (\ref{eq-Fss}), it gives
\begin{eqnarray}
[\sqrt{(\alpha+c^2)^2+2\lambda-4c^2H},S_1]&=&2c^2S_1,\\
~[\sqrt{(\alpha+c^2)^2+2\lambda-4c^2H},S_2]&=&-2c^2S_2,
\end{eqnarray}

Therefore we can define
\begin{eqnarray}
&&b_0=\frac{1}{2c^2}\sqrt{(\alpha+c^2)^2+2\lambda-4c^2 H},\\
&&b=S_2,~
b^\dagger=S_1\left[\frac{2c^2}{\sqrt{(\alpha+c^2)^2+2\lambda-4c^2
H}}-1\right].\nonumber
\end{eqnarray}
It belongs to the third case and according to the third treatment,
the Eqs. (\ref{eq-etaxi}) and Eq. (\ref{eq-case3}) becomes
\begin{eqnarray}
&\left(\frac{1}{b_0-1}-1\right)\xi(b_0-1)^2=\left(\frac{1}{b_0}-1\right)\xi(b_0)^2,\label{eq-etaxi1}&\\
& [J_+,J_-]=[S_1,S_2]\eta(b_0)\xi^2(b_0)\label{eq-J+J-}&
\end{eqnarray}
There is a simplest choice for $\xi(b_0)$ to satisfy the Eq.
(\ref{eq-etaxi1}), that is
\begin{equation}
\xi(b_0)^2=\xi_1\frac{b_0}{1-b_0},
\end{equation}
where $\xi_1$ is also a undetermined constant.
 Insert it into Eq. (\ref{eq-J+J-}), we can determine the two constants $\xi_0$ and $\xi_1$ as
\begin{equation}
\xi_0=0,\qquad \xi_1=\frac{1}{4abc^2}
\end{equation}

Finally, we can realize our algebraic structure by defining again
\begin{eqnarray}
&J_0=b_0,\quad
J_-=\frac{1}{\sqrt{4abc^2}}\sqrt{\frac{b_0}{b_0-1}}b,&\nonumber\\
&J_+= \frac{1}{\sqrt{4abc^2}}b^\dagger\sqrt{\frac{b_0}{b_0-1}}
\end{eqnarray}
which satisfy the $SU(1,1)$ algebra,
\[[J_0,J_-]=-J_-,\quad [J_0,J_+]=J_+,\quad[J_+,J_-]=-2J_0.\]

\subsection{The $X(x)=-1$, $Y(x)=a\sin(kx)+b\cos{kx}$ case}

By substituting into Eq. (\ref{eq-cosis}), we solved
\begin{eqnarray}
Q(x)&=&\frac{\lambda}{k}\left[a\cos(kx)-b\sin(kx)\right]+c_1,\\
Z(x)&=&\frac{\alpha-k^2}{2k}\left(a\cos kx-b\sin kx\right)+c_2,
\end{eqnarray}
and
\begin{equation}\label{eq-radia}
V(x)=c_3\left(a\sin kx+b\cos
kx\right)^{-2}-\frac{(\alpha-k^2)^2+2\lambda}{4k^2},
\end{equation}
with $\beta=2k^2/\lambda$, $\gamma=-2k^2c_1/\lambda$,
$\nu=-\alpha+2k^2$, $\tau=c_2(\alpha-2k^2)-2\lambda c_2$,  and
$c_1+\alpha c_2=0$. This gives us the first P\"{o}schl-Teller
potential.

The closed operator set $\{H, \tilde{Q}, \tilde{P}\}$ satisfy the
commutation relations\begin{equation} [H,
(\tilde{Q},\tilde{P})]=(\tilde{Q},\tilde{P})\left(\begin{matrix}
2k^2-\alpha&1+\frac{2k^2}{\lambda}H\\2\lambda&\alpha
\end{matrix}\right)
\end{equation}
in which
\begin{eqnarray}
\tilde{Q}=Q-c_1&=&\frac{\lambda}{k}\left(a\cos kx-b\sin kx\right)\\
\tilde{P}=P-c_2&=&\left(a\sin kx+b\cos kx\right)\frac{d}{dx}+\frac{\alpha-k^2}{2k}\nonumber\\
&&\times\left(a\cos kx-b\sin kx\right)
\end{eqnarray}
The shift operators $S_1$ and $S_2$ which satisfied the commutation
relation
\begin{eqnarray*}
[H,S_1]&=&S_1\left[k^2-\sqrt{(\alpha-k^2)^2+2\lambda+4k^2
H}\right],\\
~[H,S_2]&=&S_2\left[k^2+\sqrt{(\alpha-k^2)^2+2\lambda+4k^2
H}\right].
\end{eqnarray*}
with
\begin{eqnarray*}
S_1&=&-\tilde{Q}\frac{1}{2\lambda}\left[\alpha-k^2+\sqrt{(\alpha-k^2)^2+2\lambda+4k^2
H}\right]+\tilde{P},\nonumber\\
S_2&=&-\tilde{Q}\frac{1}{2\lambda}\left[\alpha-k^2-\sqrt{(\alpha-k^2)^2+2\lambda+4k^2
H}\right]+\tilde{P}.\nonumber\\
\end{eqnarray*}
We can derive that
\begin{eqnarray*}
S_1^\dagger=-S_2\left[1+\frac{2k^2}{\sqrt{(\alpha-k^2)^2+2\lambda+4k^2
H}}\right]
\end{eqnarray*}
 \begin{equation*}
[S_1,S_2]=-(a^2+b^2)\sqrt{(\alpha-k^2)^2+2\lambda+4k^2 H}.
\end{equation*}

From the Eq. (\ref{eq-Fss}), it gives
\begin{eqnarray}
[\sqrt{(\alpha-k^2)^2+2\lambda+4k^2
H},S_1]&=&-2k^2S_1,\\
~[\sqrt{(\alpha-k^2)^2+2\lambda+4k^2 H},S_2]&=&2k^2S_2,
\end{eqnarray}

Therefore it is easy to define
\begin{eqnarray}
&&b_0=\frac{1}{2k^2}\sqrt{(\alpha-k^2)^2+2\lambda+4k^2
H},\\
&&b=S_1,\\
&&b^\dagger=S_1^\dagger=-S_2\left[1+\frac{2k^2}{\sqrt{(\alpha-k^2)^2+2\lambda+4k^2
H}}\right].\nonumber
\end{eqnarray}
It belongs to the third case, and according to the treatment the
Eqs. (\ref{eq-etaxi}) and (\ref{eq-case3}) becomes
\begin{eqnarray}
&\left(1+\frac{1}{b_0}\right)\xi(b_0)^2=\left(1+\frac{1}{b_0-1}\right)\xi(b_0-1)^2,\label{eq-etaxi2}&\\
&[J_+,J_-]=[S_1,S_2]\left(1+\frac{1}{b_0}\right)\xi(b_0)^2.\label{eq-JJ1}&
\end{eqnarray}
The simplest choice for $\xi(b_0)$ satisfying the condition
(\ref{eq-etaxi2}) is
\begin{equation}
\xi(b_0)^2=\xi_1\frac{b_0}{b_0+1}.
\end{equation}
Taking this choice into the Eq. (\ref{eq-JJ1}), we obtain
\begin{equation}
[J_+,J_-]=-2k^2(a^2+b^2)\xi_1b_0,
\end{equation}
which at the case of the $SU(1,1)$ requires
\begin{equation}
\xi_1=\frac{1}{k^2(a^2+b^2)}
\end{equation}

Finally, we can realize our algebraic structure by defining again
\begin{eqnarray}
&J_0=b_0,\quad
J_-=\frac{1}{k\sqrt{a^2+b^2}}\sqrt{\frac{b_0}{b_0+1}}b,&\nonumber\\
&J_+= \frac{1}{k\sqrt{a^2+b^2}}b^\dagger\sqrt{\frac{b_0}{b_0+1}}
\end{eqnarray}
which satisfy the $SU(1,1)$ algebra,
\[[J_0,J_-]=-J_-,\quad [J_0,J_+]=J_+,\quad[J_+,J_-]=-2J_0.\]

\section{Extended to radial potentials}\label{sec-rad}
At case of dealing with the spherically symmetric potentials, their
wavefunctions can be divided into angular parts
$Y_{lm}(\theta,\varphi)$ and radial parts $R_l(r)$. For the radial
part $R_l(r)$, that it only depends on one variable is similar to
the one-dimensional problem, so that we can try to use the methods
introduced above to discuss it more or less.

Moreover, the radial part of the Schr\"{o}dinger equation is
generally written as
\[\left[\frac{d^2}{dr^2}+\frac{2\mu}{\hbar^2}\big(E-V(r)\big)-\frac{l(l+1)}{r^2}\right]\chi_l(r)=0,\]
where $\chi_l(r)=R_l(r)r$. We now try to discuss the radial Coulomb
problem or more its extension of Kratzer's molecular potential by
using the above method. Their Schr\"{o}dinger equations are
\begin{eqnarray*}
&\left(-\frac{d^2}{dr^2}+\frac{l(l+1)}{r^2}-\frac{2Z}{r}+\frac{Z^2}{n^2}\right)\psi_{n,l}=0&\\
\end{eqnarray*}
\begin{eqnarray*}
\Bigg[-\frac{d^2}{dr^2}&+&\frac{l(l+1)+Da^2}{r^2}-\frac{2Da}{r}\\
&+&\frac{D^2a^2}{\left(n+1/2+\sqrt{(l+1/2)^2+\gamma^2}\right)^2}\Bigg]\psi_{n,l}=0
\end{eqnarray*}
which if making the substitutions $\rho=(Z/n)r$ or
$\rho=\frac{Da}{n+1/2+\sqrt{(l+1/2)^2+\gamma^2}}r$ can be
transformed into the easily disposed forms
\begin{eqnarray*}
&\left[-\frac{d^2}{d\rho^2}+\frac{l(l+1)}{\rho^2}+1-\frac{2n}{\rho}\right]\psi_{n,l}=0,&\label{eq-hyd}\\
&\left[-\frac{d^2}{d\rho^2}+\frac{l(l+1)+Da^2}{\rho^2}+1-2\frac{n+\frac{1}{2}+\sqrt{(l+\frac{1}{2})^2+\gamma^2}}{\rho}\right]\psi_{n,l}=0.&\nonumber
\end{eqnarray*}
Further, such form has two extended deformations which by virtue of
the method discussed above are related to the energy quantum number
$n$ and the orbital quantum number $l$, respectively. One of the
deformations is
\begin{eqnarray}
\left[-\rho\frac{d^2}{d\rho^2}+\frac{l(l+1)}{\rho}+\rho\right]\psi_{n,l}=2n\psi_{n,l},\label{eq-rho1}
\end{eqnarray}
\begin{eqnarray*}
\bigg[-\rho\frac{d^2}{d\rho^2}+&&\!\!\!\!\!\!\!\!\frac{l(l+1)+Da^2}{\rho}+\rho\bigg]\psi_{n,l}=\nonumber\\
&&2\bigg[n+\frac{1}{2}+\sqrt{(l+\frac{1}{2})^2+\gamma^2}\bigg]\psi_{n,l}.
\end{eqnarray*}
And the other one is
\begin{eqnarray}
\left[-\rho^2\frac{d^2}{d\rho^2}+\rho^2-2n\rho+l^2\right]\psi_{n,l}=-l\psi_{n,l},\label{eq-rho2}
\end{eqnarray}
\begin{eqnarray*}
\bigg[-\rho^2\frac{d^2}{d\rho^2}&&\!\!\!\!\!\!\!\!+\rho^2-2\bigg(n+\frac{1}{2}+\sqrt{(l+\frac{1}{2})^2+\gamma^2}\bigg)\rho\nonumber\\
&&+l^2+Da^2\bigg]\psi_{n,l} =-l\psi_{n,l}.
\end{eqnarray*}
 The left hand of the above deformed Shr\"{o}dinger equations may
well be called pseudo-Hamiltonian, which are no longer self-adjoint.

\subsection{The $X(x)=-x$, $Y(x)=x$ case}
By substituting into Eq. (\ref{eq-cosis}), we solved
\begin{eqnarray*}
Q(x)&=&-\lambda x+c_1,\\
Z(x)&=&-\frac{\alpha}{2}x+c_2,
\end{eqnarray*}
and
\begin{equation*}
V(x)=\frac{1}{2}\left(\lambda+\frac{\alpha^2}{2}\right)x+\frac{c_3}{x}-(c_1+\alpha
c_2),
\end{equation*}
with $\beta=0$, $\gamma=1$, $\nu=-\alpha$, $\tau=-2\lambda c_2-\nu
c_1$, and $c_1+\alpha c_2=0$. This is the case of the first
deformation [see Eq. (\ref{eq-rho1})].

The closed operator set $\{H, \tilde{Q}, \tilde{P}\}$ satisfy the
commutation relations
\begin{equation*}
[H, (\tilde{Q},\tilde{P})]=(\tilde{Q},\tilde{P})\left(\begin{matrix}
-\alpha&1\\2\lambda&\alpha
\end{matrix}\right)
\end{equation*}
in which
\begin{eqnarray*}
\tilde{Q}&=&Q+\frac{2\lambda H-\alpha^2c_1+2\lambda c_2\alpha}{\alpha^2+2\lambda}\nonumber\\
&=&-\lambda x+\frac{2\lambda}{\alpha^2+2\lambda}H+2\lambda\frac{c_1+c_2\alpha}{\alpha^2+2\lambda}\\
\tilde{P}&=&P+\frac{\alpha H+\alpha c_1-2\lambda
c_2}{2\lambda+\alpha^2}\nonumber\\&=&x\frac{d}{dx}-\frac{\alpha}{2}x+\frac{\alpha}{\alpha^2+2\lambda}H+\alpha\frac{c_1+\alpha
c_2}{\alpha^2+2\lambda}
\end{eqnarray*}
The shift operators $S_1$ and $S_2$ satisfy the commutation relation
\begin{equation*}
[H,(S_1, S_2)]=(S_1,S_2)\left(\begin{matrix}
-\sqrt{\alpha^2+2\lambda}&0\\0&\sqrt{\alpha^2+2\lambda}
\end{matrix}\right).
\end{equation*}
with
\begin{eqnarray*}
S_1&=&-\tilde{Q}\frac{\alpha+\sqrt{\alpha^2+2\lambda}}{2\lambda}+\tilde{P},\nonumber\\
S_2&=&-\tilde{Q}\frac{\alpha-\sqrt{\alpha^2+2\lambda}}{2\lambda}+\tilde{P}.\nonumber\\
\end{eqnarray*}
which have
\begin{eqnarray*}
S_1^\dagger=-S_2-1+\frac{1}{\sqrt{\alpha^2+2\lambda}}\frac{d}{dx},
\end{eqnarray*}
 \begin{equation*}
[S_1,S_2]=-\frac{2}{\sqrt{\alpha^2+2\lambda}}(H+c_1+\alpha c_2).
\end{equation*}
Here if we still define
\begin{eqnarray}
b_0=\frac{H}{\sqrt{\alpha^2+2\lambda}},\quad b_-=S_1,\quad
b_+=S_2.\nonumber
\end{eqnarray}
they can satisfy
\begin{equation*}
[b_0,b]=-b,\quad [b_0,b^\dagger]=b^\dagger
\end{equation*}
but $b_0^\dagger\neq b_0$ and $b_-^\dagger\neq b_+$. So for such
pseudo-Hamiltonian we can realize its algebraic structure by
defining again
\begin{equation}
J_0=b_0+\frac{c_1+\alpha c_2}{\sqrt{\alpha^2+2\lambda}},\quad
J_+=b^\dagger,\quad J_-=b.
\end{equation}
which satisfy $[J_+,J_-]=-2J_0$ but at a price of $J_0^\dagger\neq
J_0$ and $J_-^\dagger\neq J_+$. So it does not mean we have
constructed a SU(1,1) algebra.

\subsection{The $X(x)=-x^2$, $Y(x)=1$ case}

By substituting into Eq. (\ref{eq-cosis}), we solved
\begin{eqnarray}
Q(x)&=&\frac{\lambda}{x}+\frac{\tau}{2},\\
Z(x)&=&0,
\end{eqnarray}
and
\begin{equation*}
V(x)=-c_3x^2+\frac{\gamma\lambda}{2}x+\frac{\lambda}{2}
\end{equation*}
with $\beta=-2/\lambda$, $2\gamma=-\beta\tau$, $\nu=-2$, and
$\alpha=0$. This is simply the case of the second deformation [see
Eq. (\ref{eq-rho2})].

The closed operator set $\{H, \tilde{Q}, \tilde{P}\}$ satisfy the
commutation relation
\begin{equation*}
[H, (\tilde{Q},\tilde{P})]=(\tilde{Q},\tilde{P})\left(\begin{matrix}
-2&-\frac{2}{\lambda}H+1 \\2\lambda&0
\end{matrix}\right)
\end{equation*}
in which
\begin{eqnarray*}
\tilde{Q}&=&\frac{\lambda}{x}+\frac{\tau}{2}+\frac{\tau}{\lambda-2H}H,\\
\tilde{P}&=&\frac{d}{dx}+\frac{\tau}{2(\lambda-2H)}.
\end{eqnarray*}
\begin{eqnarray*}
&&[H,(S_1, S_2)]=(S_1,S_2)\\&&\left(\begin{matrix}
-1+\sqrt{1+2\lambda-4H}&0\\0&-1-\sqrt{1+2\lambda-4H}
\end{matrix}\right).
\end{eqnarray*}
\begin{eqnarray*}
&&\left[\sqrt{1+2\lambda-4H},(S_1,
S_2)\right]=(S_1,S_2)\left(\begin{matrix} -2&0\\0&2
\end{matrix}\right).
\end{eqnarray*}
with
\begin{eqnarray*}
S_1&=&\tilde{Q}\left(\sqrt{1+2\lambda-4H}-1\right)+2\lambda\tilde{P},\nonumber\\
S_2&=&-\tilde{Q}\left(\sqrt{1+2\lambda-4H}+1\right)+2\lambda\tilde{P}.\nonumber\\
\end{eqnarray*}
So we can define \begin{equation*}
J_0=\frac{1}{2}\sqrt{1+2\lambda-4H},\  J_-=S_1,\ J_+=S_2
\end{equation*}
which can satisfy
\begin{equation*}
[J_0,J_-]=-J_-,\quad [J_0,J_+]=J_+
\end{equation*}
\section{discussion and conclusion}
We mainly explore the linear algebraic structure like $SU(2)$ or
$SU(1,1)$ of the shift operators for some one-dimensional exactly
solvable potentials in this paper. During such process, a set of
method based on original diagonalizing technique is presented to
construct those suitable operator elements, $J_0$, $J_\pm$ that
satisfy $SU(2)$ or $SU(1,1)$ algebra. A quick glance at the energy
levels of the various cases shows that new-defined element operator
$J_0$ has the same eigenvalues as that of the Hamiltonian of the
harmonic oscillator. This fact also confirms to some extent that the
local behavior of the most solvable potentials reduces to the
harmonic oscillator \cite{Wehrhahn(1992)}. With $J_-\psi_0=0$, we
can get the ground state $\psi_0$, and $J_+\psi_n=C\psi_{n+1}$ to
get the whole spectrum. At the same time, $J_0\psi_n=n\psi_n$ would
indirectly give out the eigenvalues of Hamiltonian $H$.

Since the importance of spherically symmetric potentials in quantum
mechanics, in Sec. \ref{sec-rad} we discuss the deformed radial
Hamiltonian of the hydrogen atom and Kratzer's molecular potential,
though they do not have a complete $SU(1,1)$ algebra for the
non-hermite pseudo-Hamiltonian. By exploring the un-deformed
Hamiltonian with known radial raising and lowering operators
\cite{Newmarch(1978),Chen1(2000)}, it is expected that a complete
$SU(1,1)$ algebra may arises.

At last, the similarity between radial problem and one-dimensional
potentials encourages us to deal with the radial problem in the same
way. And the corresponding algebra turns to approach $SU(1,1)$
algebra but for $J_0\neq J_0^\dagger$, $J_+^\dagger\neq J_-$.

\begin{acknowledgments}

\end{acknowledgments}

\end{document}